\PassOptionsToPackage{unicode}{hyperref}
\PassOptionsToPackage{hyphens}{url}
\PassOptionsToPackage{dvipsnames,svgnames,x11names}{xcolor}
\documentclass[
  12pt]{article}

\usepackage{amsmath,amssymb}
\usepackage{iftex}
\ifPDFTeX
  \usepackage[T1]{fontenc}
  \usepackage[utf8]{inputenc}
  \usepackage{textcomp} 
\else 
  \usepackage{unicode-math}
  \defaultfontfeatures{Scale=MatchLowercase}
  \defaultfontfeatures[\rmfamily]{Ligatures=TeX,Scale=1}
\fi
\usepackage{lmodern}

\usepackage{amsmath}
\usepackage{amssymb}
\usepackage{dsfont}
\usepackage{natbib}
\setcitestyle{authoryear,open={(},close={)}}
\usepackage{hyperref}
\usepackage{xcolor}
\usepackage[ruled, vlined]{algorithm2e}
\usepackage{graphicx}
\usepackage{multirow}

\usepackage{soul}

\definecolor{eGreen}{rgb}{.057, .549,.065}

\newcommand{\btheta}{\boldsymbol{\theta}}
\newcommand{\bmu}{\boldsymbol{\mu}}
\newcommand{\br}{\mathbf{r}}
\newcommand{\bi}{\mathbf{i}}
\newcommand{\bz}{\mathbf{z}}
\newcommand{\bZ}{\mathbf{Z}}

\newcommand{\bY}{\mathbf{Y}}

\newcommand{\Rmax}{\mathcal{R}_\text{max}}
\newcommand{\cR}{\mathcal{R}}
\newcommand{\cZ}{\mathcal{Z}}
\newcommand{\cS}{\mathcal{S}}
\newcommand{\tf}{\tilde{f}}

\ifPDFTeX\else  
\fi
\IfFileExists{upquote.sty}{\usepackage{upquote}}{}
\IfFileExists{microtype.sty}{
  \usepackage[]{microtype}
  \UseMicrotypeSet[protrusion]{basicmath} 
}{}
\makeatletter
\@ifundefined{KOMAClassName}{
  \IfFileExists{parskip.sty}{%
    \usepackage{parskip}
  }{
    \setlength{\parindent}{0pt}
    \setlength{\parskip}{6pt plus 2pt minus 1pt}}
}{
  \KOMAoptions{parskip=half}}
\makeatother
\usepackage{xcolor}
\makeatletter
\ifx\paragraph\undefined\else
  \let\oldparagraph\paragraph
  \renewcommand{\paragraph}{
    \@ifstar
      \xxxParagraphStar
      \xxxParagraphNoStar
  }
  \newcommand{\xxxParagraphStar}[1]{\oldparagraph*{#1}\mbox{}}
  \newcommand{\xxxParagraphNoStar}[1]{\oldparagraph{#1}\mbox{}}
\fi
\ifx\subparagraph\undefined\else
  \let\oldsubparagraph\subparagraph
  \renewcommand{\subparagraph}{
    \@ifstar
      \xxxSubParagraphStar
      \xxxSubParagraphNoStar
  }
  \newcommand{\xxxSubParagraphStar}[1]{\oldsubparagraph*{#1}\mbox{}}
  \newcommand{\xxxSubParagraphNoStar}[1]{\oldsubparagraph{#1}\mbox{}}
\fi
\makeatother

\usepackage{longtable,booktabs,array}
\usepackage{calc} 
\usepackage{etoolbox}
\makeatletter
\patchcmd\longtable{\par}{\if@noskipsec\mbox{}\fi\par}{}{}
\makeatother
\IfFileExists{footnotehyper.sty}{\usepackage{footnotehyper}}{\usepackage{footnote}}
\makesavenoteenv{longtable}
\usepackage{graphicx}
\makeatletter
\def\maxwidth{\ifdim\Gin@nat@width>\linewidth\linewidth\else\Gin@nat@width\fi}
\def\maxheight{\ifdim\Gin@nat@height>\textheight\textheight\else\Gin@nat@height\fi}
\makeatother
\setkeys{Gin}{width=\maxwidth,height=\maxheight,keepaspectratio}
\makeatletter
\def\fps@figure{htbp}
\makeatother

\makeatletter
\@ifpackageloaded{caption}{}{\usepackage{caption}}
\AtBeginDocument{%
\ifdefined\contentsname
  \renewcommand*\contentsname{Table of contents}
\else
  \newcommand\contentsname{Table of contents}
\fi
\ifdefined\listfigurename
  \renewcommand*\listfigurename{List of Figures}
\else
  \newcommand\listfigurename{List of Figures}
\fi
\ifdefined\listtablename
  \renewcommand*\listtablename{List of Tables}
\else
  \newcommand\listtablename{List of Tables}
\fi
\ifdefined\figurename
  \renewcommand*\figurename{Figure}
\else
  \newcommand\figurename{Figure}
\fi
\ifdefined\tablename
  \renewcommand*\tablename{Table}
\else
  \newcommand\tablename{Table}
\fi
}
\@ifpackageloaded{float}{}{\usepackage{float}}
\floatstyle{ruled}
\@ifundefined{c@chapter}{\newfloat{codelisting}{h}{lop}}{\newfloat{codelisting}{h}{lop}[chapter]}
\floatname{codelisting}{Listing}

\makeatother
\makeatletter
\@ifpackageloaded{caption}{}{\usepackage{caption}}
\@ifpackageloaded{subcaption}{}{\usepackage{subcaption}}
\makeatother

\ifLuaTeX
  \usepackage{selnolig}  
\fi
\usepackage[]{natbib}
\usepackage{bookmark}

\IfFileExists{xurl.sty}{\usepackage{xurl}}{} 
\hypersetup{
  pdftitle={Modeling Interaction Structure in High-Throughput Phenotyping in Tropical Forage Breeding Using Hidden Sparse Markov Random Fields},
  pdfkeywords={Reversible jump Markov chain Monte Carlo; Pseudoposteriors; Pseudolikelihood},
  colorlinks=true,
  linkcolor={blue},
  filecolor={Maroon},
  citecolor={Blue},
  urlcolor={Blue},
  pdfcreator={LaTeX via pandoc}}

\newcommand{\anon}{1}

\begin{document}

\def\spacingset#1{\renewcommand{\baselinestretch}%
{#1}\small\normalsize} \spacingset{1}


\if1\anon
{
  \title{\bf Modeling Interaction Structure in High-Throughput Phenotyping in Tropical Forage Breeding Using Hidden Sparse Markov Random Fields}
  \author{
    Victor Freguglia$^{1}$, Nancy Lopes Garcia$^{1}$, \\
    Guilherme Francio Niederauer$^{2}$, Mateus Figueiredo$^{3}$ \\
    \small{$^{1}$ Department of Statistics, Universidade Estadual de Campinas} \\
    \small{$^{2}$ Department of Plant Biology, Universidade Estadual de Campinas} \\
    \small{$^{3}$ Embrapa}}
  \maketitle
} \fi

\if0\anon
{
  \bigskip
  \bigskip
  \bigskip
  \begin{center}
    {\LARGE\bf Modeling Interaction Structure in High-Throughput Phenotyping in Tropical Forage Breeding Using Hidden Sparse Markov Random Fields}
\end{center}
  \medskip
} \fi

\bigskip
\begin{abstract}
High-throughput phenotyping, combined with remote sensing using images captured by drones, can be a powerful tool for monitoring and analyzing phenotypic traits in plants and agricultural crops, especially forage species. From high-resolution aerial images, it is possible to use statistical image-processing techniques to obtain precise data on plant morphology. The use of drones enables large-scale data collection in a fast and efficient manner, without the need for invasive field interventions. However, advanced image-processing techniques are required to extract detailed information, providing insights into crop performance and supporting the selection of varieties that are better adapted to the environment or more resistant to diseases and pests. A Gaussian mixture driven by a hidden Markov random field (MRF) is employed to model these images in order to separate the plants from the background  from the images that enable the digital definition of different phenotypic measurements. One challenge is the problem of estimating the interacting neighborhood of the hidden MRF and the components of the Gaussian mixture.  Using a Bayesian framework, we propose a Reversible Jump Monte Carlo Markov Chain algorithm that jumps across subsets of a maximal range neighborhood, allowing us to perform model selection based on a marginal pseudoposterior distribution. 
\end{abstract}

\noindent%
\vfill

\newpage
\spacingset{1.8} 

\section{Introduction} 

Many vegetation segmentation methods rely on transforming the original image using a vegetation index or a similar measure, followed by thresholding to distinguish plant pixels from soil. Although straightforward and computationally efficient, these approaches can be sensitive to variations in plant appearance, particularly when parts of the vegetation have colors similar to the background. 

Machine learning and deep learning approaches have also been increasingly used for vegetation segmentation, often providing greater flexibility in handling complex images. However, these methods may require substantial amounts of annotated training data, considerable computational resources, and specialized expertise for model development and deployment. These requirements can limit their accessibility in practical breeding programs and on-farm applications, where resources and technical support may be limited. Consequently, there remains a need for accessible methods that can account for spatial dependence in images while avoiding the computational and data requirements of more complex learning approaches. The proposed MRF framework addresses this need by incorporating spatial relationships between neighboring pixels and allowing the interaction structure to be inferred from the observed image.

\subsection{High-Throughput Phenotyping (HTP)} \label{sec:motivating}

Tropical forage species play a crucial role in livestock production systems, underpinning a large share of Brazil's agribusiness sector, which moved approximately R\$913.14 billion in 2021 \citep{abiec2026}. Among the main forage varieties used for pasture in Brazil, those derived from \textit{Megathyrsus maximus} stand out, as this species is considered the most productive among tropical forages \citep{valle2009melhoramento}. However, the growing demand for animal protein, combined with the intensifying effects of climate change, has increased the need to select better adapted and more productive plant materials.

Despite their importance, breeding programs for these crops remain relatively limited in efficiency and still offer substantial room for improvement, largely due to the constraints of conventional phenotyping methods. These methods rely heavily on manual measurements, which are labor-intensive, subjective, and difficult to scale \citep{araus_field_2014, Singh2016}. Moreover, they are often unsuitable for capturing the full range of phenotypic traits relevant to genetic breeding, resulting in incomplete and imprecise data collection and, consequently, longer timelines for selecting new varieties \citep{furbank_phenomics_2011}. These challenges highlight the need for new field management and phenotyping strategies capable of overcoming such limitations.

High-throughput phenotyping (HTP), particularly when combined with drone-based remote sensing, has strong potential to significantly enhance breeding programs by enabling rapid, large-scale, precise, and non-destructive assessment of complex agronomic traits. Drone-acquired images can be used to extract morphological and physiological characteristics, enabling the quantification of a wide range of phenotypic measurements. In particular, two types of sensors are commonly used:

\begin{itemize}
    \item \textbf{RGB sensor} (red, green, blue), for detecting visible wavelengths. This is a more accessible sensor that can be used to estimate vegetation indices, plant height, plant structure, growth rates, and morphological characteristics.
    \item \textbf{NIR sensor} (near-infrared), used to detect the higher reflectance of green plant areas in the 700--1300~nm range, which can be used to determine chlorophyll content, water status, and vegetation indices.
\end{itemize}

\begin{figure}[ht!]
\centering
\includegraphics[width=1\linewidth]{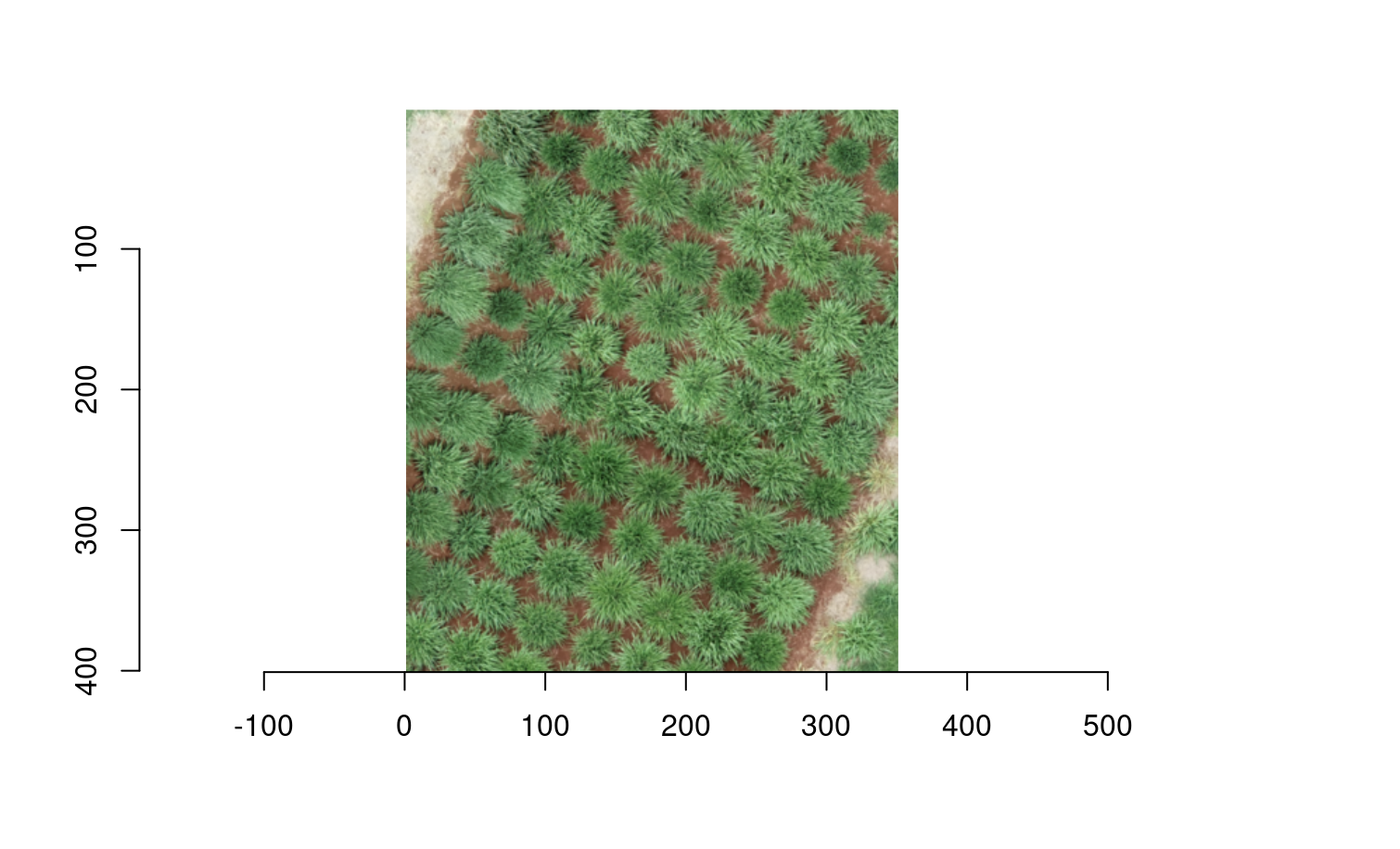}
\caption{Image captured by RGB sensor.} \label{fig:planta}
\end{figure}

An example of an image used for analysis is shown in Figure~\ref{fig:planta}. Based on these images, vegetation indices are computed to quantify spectral characteristics associated with with agronomically relevant traits. Several vegetation indices proposed in the literature. In this work, we use the so called Excess Green  ($ExG = 2G - R - B$) to isolate green plant tissue from background soil, crop residue, and varying lighting conditions in digital color images as proposed by \citep{woebbecke1995color}. These indices are computed on a pixel-by-pixel basis across the image.


With the increasing volume of data generated by these acquisition methods, new approaches for processing and analyzing the resulting imagery are also required. One of the main challenges in remote sensing imagery is discriminating pixels of interest (i.e., plant material) from background elements such as soil and other non-plant objects. In addition to this pixel-level segmentation, identifying individual plants within a crop field can be equally challenging, as neighboring plants frequently overlap, making it difficult to isolate and associate individual plants in an image with their corresponding ground-truth data.

\subsection{Markov random fields (MRFs)}

Markov Random Fields on two-dimensional lattices are popular
probabilistic models for describing features of
digital images in a wide range of applications.
Classical problems like image segmentation rely on these models to describe unobserved
variables used for pixel classification
\citep{held1997markov, zhang2001segmentation},
while more general inference-oriented models describe pixel values directly as a MRF,
for example, in texture modeling problems \citep{hassner1981use, cross1983markov,freguglia2020hidden}.

One of the main inferential challenges in MRFs modeling
is the fact that their undelying dependence structure depends on a graph which contains cycles that prevents factorizing the joint distribution into simple conditional probabilities as in acyclic Markov chains. Consequently, calculating the parameter-dependent normalizing constant (or partition function) requires evaluating a high-dimensional sum or integral over an exponentially large state space.  Whenever this high-dimensional integral cannot be computed in reasonable time,the likelihood function becomes intractable, rendering standard likelihood-based inference and model selection methods inapplicable.

A common strategy replaces the likelihood function in the posterior distribution with the pseudolikelihood function introduced by \citet{besag1975statistical}, yielding an inference framework based on what is known as the pseudoposterior distribution rather than the exact posterior. Although the pseudolikelihood function may differ from the actual likelihood in key properties (such as its mode), pseudolikelihood-based methods enjoy theoretical guarantees and demonstrated practical utility in Bayesian settings—often incorporating calibration adjustments such as those proposed by \citet{bouranis2017efficient}. Pseudolikelihoods have also been widely adopted to identify the neighborhood structure in Markov random fields (MRFs).
In particular, \citet{csiszar2006consistent} proved that a modification of the Bayesian Information Criteria, replacing
the likelihood by the pseudolikelihood, provides strongly consistent estimators of the basic neighborhood
from a single realization of the process observed over increasing regions.
Under the Bayesian paradigm, marginal pseudoposterior likelihood has been used to find the dependence structure in Markov networks \citep[see][and references therein]{pensar2017marginal}.

Model selection remains a critical challenge in Markov random field (MRF) modeling,where standard methodologies are often rendered inapplicable by intractable likelihoods. Furthermore, using an approximation of the likelihood function removes most of the probabilistic properties that is the framework for 
model selection and hypothesis testing. One advantage of studying complex models under a Bayesian framework is that the space of unobserved
random quantities may be extended to include not only a vector of real-valued parameters, but also abstract objects indexing models, such as neighborhoods.
Sampling from these general objects using MCMC methods
tends to be more feasible than constructing efficient optimizations algorithm in such arbitrary spaces.
In this context, Reversible Jump Monte Carlo Markov Chain (RJMCMC) algorithms,  \citep{green1995reversible} 
provide a framework for constructing a Markov Chain with specific invariant distribution on general spaces, including varying-dimensional parameter spaces that are highly useful for
variable and model selection under a Bayesian philosophy. To tackle intractable likelihoods during Bayesian model selection, recent strategies pair RJMCMC with pseudolikelihood approximations. For instance, \citet{arnesen2017prior} implemented this approach to select the dependence structure of Ising models among low-dimensional candidate spaces, while \citet{BOURANIS2018221} developed a pseudolikelihood-based RJMCMC procedure for model selection in exponential random graph models. Although RJMCMC is well suited for Bayesian model choice, achieving reasonable mixing and computational efficiency requires a carefully engineered proposal kernel.  

\subsection{Our approach}

In this work, we develop a methodology for extracting relevant features
from RGB images such as the example shown in Figure~\ref{fig:planta}.
The proposed framework is based on a hierarchical mixture model driven
by a latent Markov random field (MRF) with an unknown interaction
structure, as described in \autoref{sec:model}. We introduce a pseudoposterior-based approach for
selecting the interaction structure from a large collection of candidate
subsets of a maximal neighborhood system. The methodology applies to a
general class of pairwise-interaction MRF models defined through sets of
relative positions, as introduced in Subsection \ref{sec:mrf}. Inference is conducted
within a Bayesian framework using a Gibbs sampler augmented with
Metropolis--Hastings updates and Reversible Jump Markov Chain Monte
Carlo (RJMCMC) moves to infer the underlying neighborhood system as described in Section \ref{sec:bayesian}.  Model selection
is performed via RJMCMC using a transition kernel specifically designed
for efficient exploration of the space of admissible interaction
structures, as detailed in Subsection \ref{sec:RJMCMC}. The performance of the proposed methodology is assessed through an
extensive simulation study under a variety of scenarios
(\autoref{sec:simulation}). Finally, in \autoref{sec:application}, we
demonstrate its practical utility in the data described in Subsection \ref{sec:motivating}.

\section{The model} \label{sec:model}

Hidden Markov random fields (HMRFs) have been applied for color-based image segmentation, for example in \citet{barker2000unsupervised} using a Bayesian approach, and in \citet{zhang2001segmentation}, where an EM algorithm is proposed for segmentation and parameter estimation assuming a fixed interaction structure in the field. Our objective is to adopt a similar approach to obtain a coarse-scale characterization of the image, treating it as a hidden mean color plus random noise to account for pixel values on a continuous scale. In our case, however, we face the additional challenge that nearest-neighbor interactions may not adequately capture the intricate features of our images, and thus the interaction structure itself must be estimated.

\subsection{Gaussian Mixture model} \label{sec:gaussian}

	

To account for spatial dependence among the pixel values $\mathbf{Y}$ defined on a rectangular lattice $\mathcal{S}$ (see Figure~\ref{fig:planta}), we decompose the image into a coarse-scale spatial component driven by a latent Markov random field $\mathbf{Z}$ and an observational error vector $\boldsymbol{\epsilon}$. In this way, we obtain an estimation method that is as flexible as mixture models for fitting the data while also exploiting spatial information at a coarse scale. Specifically, we assume a hidden structure comprising a finite set of colors $\{\mu_1, \ldots, \mu_C\}$. For each pixel $i$, given its latent class $z_i \in \{1, \ldots, C\}$, the observed color is modeled as
 $$y_i = \mu_{z_i} + \epsilon_i,$$
 where $\boldsymbol{\epsilon} = \{\epsilon_i : i \in \cS \}$ is an array of independent $N(0,\sigma^2)$ random variables. The field $\mathbf{Z}$ is latent and therefore unobserved. Although, for simplicity, we assume a common variance $\sigma^2$, the model can be naturally extended to allow variances that depend on $\mathbf{Z}$.
		
Note that directly modeling images with Markov random fields is computationally infeasible because (as we will see in Section \ref{sec:mrf}) the number of parameters in the model is of order $C^2$. In order to account for finer scales, our strategy is to consider pixel colors as continuous variables governed by a field $\bmu$ driven by a hidden Markov field $\mathbf{Z}$ plus an observational error.  The coarse-scale components represent unobserved physical classes $\{1, \ldots, C\}$ (such as vegetation, soil, or shadows) that index the parameter vector $\boldsymbol{\mu} = (\mu_1, \dots, \mu_C)^\top$. Therefore, the conditional distribution of observed values $\mathbf{Y}$ given a realization of the underlying field $\mathbf{Z}= \mathbf{z}$ is given by
	\begin{equation}\label{eq_mixture_cond}
	L(\mathbf{y}|\mathbf{Z}= \mathbf{z}, \bmu, \sigma^2) = \prod_{i \in \mathcal{L}} \frac{1}{\sqrt{2\pi\sigma^2}} \exp{\left({- \frac{(y_i -  \mu_{z_i})^2}{2\sigma^2}}\right)}.
	\end{equation}

\subsection{Markov Random Fields with Spatially Homogeneous Pairwise Interactions}\label{sec:mrf}


In this manuscript we consider the Markov Random Field (MRF) model on two-dimensional lattices with finite support and
non-parametric pairwise interactions as described in \citet{freguglia2020hidden} (see references therein). 
The probability function for this model is completely defined by two main elements:
\textit{a set of relative positions}, that described the interaction structure of the process,
and \textit{a vector of potentials} describing the weights of interactions for each of these relative positions.

We denote by $\cS$ a set of sites (also referred as pixels) in a finite $n_1$ by $n_2$ two-dimensional lattice
( $ \cS = \{\bi = (i_1, i_2) : 1 \leq i_1 \leq n_1, 1 \leq i_2 \leq n_2 \}$),
and $\bZ = (Z_\bi)_{\bi \in \cS}$ a random field indexed by $\cS$,
where each $Z_\bi$ is a random variable assuming values in a finite alphabet denoted $\cZ = \{0, 1, \dots, C\}$.

We define a {\it Relative Position Set (RPS)}, denoted $\cR$, as a finite set of integer vectors $\br \in \mathbb{Z}^2$ without pairs of vectors with opposing directions ($\br \in \cR \implies -\br \not\in \cR,$)
and, given a fixed RPS $\cR$, we define a vector of potentials denoted $\btheta$, as a vector of real numbers indexed by $\cZ \times \cZ \times \cR$,
$$\btheta = (\theta_{a,b,\br})_{a,b \in \cZ, \br \in \cR}.$$

Given a RPS $\cR$ and an associated vector $\btheta$,
the Markov Random Field with homogeneous pairwise interaction considered in this work is characterized by the probability measure
\begin{equation}\label{eq:mrf_likelihood}
  f(\bz|\cR, \btheta) = \frac{1}{\zeta(\btheta)} 
  \exp\left(
    \sum_{\bi \in \cS}
    \sum_{\br \in \cR}
    \sum_{a = 0}^C
    \sum_{b = 0}^C
    \theta_{a, b, \br}\mathds{1}_{(z_\bi = a)}\mathds{1}_{(z_{\bi + \br} = b)}
  \right),
\end{equation}
where $\zeta(\btheta) = \sum\limits_{\bz' \in \cZ^{|\cS|}} \exp \left( \sum\limits_{\bi \in \cS} \sum\limits_{\br \in \cR} 
    \sum\limits_{a = 0}^C
    \sum\limits_{b = 0}^C
    \theta_{a, b, \br}\mathds{1}_{(z_\bi' = a)}\mathds{1}_{(z_{\bi + \br}' = b)}
\right)$ is a normalizing constant,
with the convention that the term $\mathds{1}_{(z_\bi' = a)}$ is treated as $0$ if $\bi' \not\in \cS$ for every $a \in \cZ$. 
This ensures that the sum terms are consistently defined across every pair of positions in $\cS$ that are within a relative position in $\cR$.

In order to obtain identifiability, additional constraints are required and, following
\citet{freguglia2020hidden}, we adopt the zero-valued reference pair ($(a,b) = (0,0)$) constraint
\begin{equation*}
  \theta_{0,0, \br} = 0 \text{ for all } \br \in \cR.
\end{equation*}
Note that, while we still use the term $\theta_{0,0,\br}$ in some equations, for simplicity of notation,
we will not consider these indexes in the vector $\btheta$.
Additionally, the vector $\btheta$ can be expressed in terms of subvectors $\btheta = (\btheta_\br)_{\br \in \cR}$,
where each subvector $\btheta_\br = (\theta_{a,b,\br})_{(a,b) \in \cZ^2, (a,b) \neq (0,0)}$
corresponds to non-null potentials associated with a single relative position $\br$
and we denote by $d$ the dimension of $\btheta_\br$, which is given by $d = (|\cZ|)^2 - 1$.

\paragraph*{Conditional probabilities and Pseudolikelihood}

While the MRF model introduced is well-defined, inference for such model
gets problematic on non-trivial cases due to the intractability of the normalizing constant,
$\zeta(\btheta)$,
as it requires computing a sum of an exponential number of terms, $(|\cZ|)^{n_1n_2}$,
which quickly becomes infeasible.

One of the most important features of MRF models is local dependence that
makes probability functions decomposable into a product of
functions that depend on $\bz$ only through subsets of it, like pairs $(z_\bi, z_{\bi + \br})$,
in the case of \eqref{eq:mrf_likelihood}.
This decomposition allows expressing the conditional probability of specific $z_\bi$
given every other element $\bz_{-\bi} = \{z_{\bi'}: \bi' \in \cS, \bi' \neq \bi\}$ as
\begin{equation}\label{eq:conditional_probs}
  f(z_\bi | \bz_{-\bi}, \cR, \btheta) =
  f(z_\bi | \bz_{\mathcal{N}_\bi}, \cR, \btheta) =
  \frac{\exp \left( \sum\limits_{\br \in \cR} \theta_{z_\bi, z_{\bi + \br}, \br} + \theta_{z_{\bi - \br}, z_i, \br} \right)
  }{
    \sum\limits_{a \in \cZ}
    \exp \left( \sum\limits_{\br \in \cR} \theta_{a, z_{\bi + \br}, \br} + \theta_{z_{\bi - \br}, a, \br} \right)
},
\end{equation}
where $\mathcal{N}_\bi \subset \cS$ denotes the set of neighbors of $\bi$ based on the RPS $\cR$,
i.e.,
$\mathcal{N}_\bi = \{\bi' : \bi' \in \cS \text{ and }\bi' = \bi \pm \br, \br \in \cR\}$.

The computationally simple expressions for conditional probabilities on \eqref{eq:conditional_probs} allows the
use of alternative functions based on conditional probabilities instead of the joint probability. For problems with high-dimensional dependent data, 
when conditional probabilities are available and simple, a
function widely used as a proxy for the likelihood function 
is the {\it pseudolikelihood} function \citep{besag1975statistical}, 
defined as the product of conditional probabilities evaluated at the observed values, $z_\bi$,
\begin{equation}\label{eq:mrf_pseudolik}
  \tf(\bz | \cR, \btheta) = \prod_{\bi \in \cS} f(z_\bi | \bz_{\mathcal{N}_\bi}, \mathcal{R}, \btheta).
\end{equation}

Note that while the normalizing constant $\zeta(\btheta)$ from \eqref{eq:mrf_likelihood} requires a sum over
$(|\cZ|)^{|\cS|}$ random field configurations, \eqref{eq:mrf_pseudolik} involves
$|\cS|$ normalizing constants that are sums over $|\cZ|$ terms.
Thus, the computational cost for evaluating the pseudolikelihood is
$\mathcal{O}\left(|\cZ| \times |\cS| \right)$,
while the exact likelihood function has a cost of order $\mathcal{O}\left(|\cZ|^{|\cS|}\right)$.


\section{Inference procedure}\label{sec:bayesian}

A formal Bayesian analysis of mixture models often yields intractable calculations. To address this, data augmentation serves as an efficient procedure, enabling feasible computation via Gibbs sampling \citep{diebolt1994estimation}. Under this framework, unobservable quantities,  such as the model parameters and the latent field, are treated as random variables governed by distinct prior distributions. Many Bayesian model selection methodologies extend this paradigm by assuming that both the real-valued parameters (the vector of free potentials $\boldsymbol{\theta}$ within the scope of this work) and the model itself (interpreted as the Random Pairwise Structure, or RPS $\mathcal{R}$) are unobserved random objects with assigned priors. Since the support of $\mathcal{R}$ represents the set of interacting positions, a natural choice for the collection of candidate models $\mathcal{M}$ is the power set of a maximal RPS, denoted as $\mathcal{R}_{\text{max}}$:$$\mathcal{M} = \{\mathcal{R}': \mathcal{R}' \subset \mathcal{R}_{\text{max}}\},$$which encompasses $2^{|\mathcal{R}_{\text{max}}|}$ possible neighborhoods. For notational simplicity, we use $\boldsymbol{\theta}$ to denote a vector of varying dimension whose indexing is inherently tied to the interaction structure $\mathcal{R}$. The dimension $d_{|\mathcal{R}|}$ and the indexing of $\boldsymbol{\theta} = (\boldsymbol{\theta}_{\mathbf{r}})_{\mathbf{r} \in \mathcal{R}}$ are always implicitly specified, as the vector is consistently matched with its corresponding interaction structure $\mathcal{R}$ across all expressions. In this context, we explicitly refer to the RPS that defines the interaction structure of a Markov Random Field (MRF) as ``the model''. Consequently, this formulation can be interpreted as a model selection problem: any vector of interaction coefficients associated with an RPS $\mathcal{R}$, subject to the restriction that $\boldsymbol{\theta}_{a,b,\mathbf{r}} = 0$ for all $a,b$ and a specific $\mathbf{r}$, is likelihood-equivalent to a model that excludes $\mathbf{r}$ from $\mathcal{R}$.

\subsection{Prior specification}

Let $\boldsymbol{\Theta} = (\boldsymbol{\mu}, \sigma^2, \mathbf{z}, \boldsymbol{\theta}, \mathcal{R})$ denote the augmented parameter vector of the model, where the latent field $\mathbf{z}$ is regarded as an augmented parameter. Given a collection of valid RPSs  $\mathcal{M}$, the hierarchical Bayesian framework is characterized by the following prior specification:
$$\begin{aligned}
\boldsymbol{\mu} &\sim  N(\mathbf{0}, \tau^2 \mathbf{I}), \quad \boldsymbol{\mu} \in \mathbb{R}^{C}, \\
\sigma^2 &\sim  \text{IG}(a,b), \quad \sigma^2 > 0, \\
\mathcal{R} &\sim q(\mathcal{R}) =  \frac{\beta^{-\alpha d|\cR|}}{\eta_\mathcal{M}(\alpha,\beta)}, \quad \mathcal{R} \in \mathcal{M}, \\
\boldsymbol{\theta} \mid \mathcal{R} &\sim \phi(\boldsymbol{\theta} \mid \mathcal{R}) = 
  \frac{1}{(2\pi\sigma^2_p)^{|\cR|d/2}}
  \exp \left(
    -\frac{1}{2\sigma^2_p} \sum_{\br \in \cR} \btheta_{\br}^\top \btheta_{\br}
  \right), \quad \boldsymbol{\theta} \in \mathbb{R}^{d_{|\mathcal{R}|}},
\end{aligned}$$and$$\mathbf{z} \mid \mathcal{R}, \boldsymbol{\theta} \sim f(\mathbf{z} \mid \mathcal{R}, \boldsymbol{\theta}), \quad \mathbf{z} \in \mathcal{Z}^{|\mathcal{S}|},$$
where $f(\mathbf{z} \mid \mathcal{R}, \boldsymbol{\theta})$ represents the probability mass function of the latent MRF defined in \eqref{eq:mrf_likelihood}.  The normalizing constant $\eta_\mathcal{M}(\alpha,\beta) = \sum_{\cR' \in \mathcal{M}} \beta^{-\alpha d|\cR|}$ is makes $q$ a  proper distribution on $\mathcal{M}$.

\subsection{Full Conditional posterior distributions}

Following the Bayesian paradigm, the posterior distribution
$p(\boldsymbol{\Theta} \mid \mathbf{y})$ is proportional to the product
of the likelihood function and the prior distribution of
$\boldsymbol{\Theta}$. Since the resulting posterior distribution does
not admit a closed-form expression, inference is carried out using
Markov chain Monte Carlo (MCMC) methods. In particular, we employ a
Gibbs sampler augmented with Reversible Jump MCMC steps to generate
samples from the target posterior distribution, see
\citet{green1995reversible} and \citet{gamerman2006markov} for further
details.

\subsubsection{Full Conditional posterior distribution of $\boldsymbol{\mu}$}

\begin{eqnarray} \label{eq_cond_mu}
\lefteqn{P(\boldsymbol{\mu} \mid \mathbf{y}, \sigma^2, \mathbf{z}, \boldsymbol{\theta}, \mathcal{R}) = P(\boldsymbol{\mu} \mid \mathbf{y}, \sigma^2, \mathbf{z}) } \\
&&\propto \frac{1}{\left(\sqrt{2 \pi \tau^2}\right)^C} \exp\left\{-\frac{1}{2\tau^2} \sum_{j=1}^{C} \mu_j^2\right\}  
\prod_{i \in \mathcal{L}} \frac{1}{\sqrt{2 \pi \sigma^2}} \exp\left\{-\frac{(y_i - \mu_{z_i})^2}{2 \sigma^2}\right\}. \nonumber
\end{eqnarray}

Therefore, each $\mu_j$ is independent normally distributed with mean \\
$ \frac{n_j}{\sigma^2}(\frac{1}{\tau^2} + \frac{n_j}{\sigma^2})^{-1} \bar{y_j}$  and variance 
$(\frac{1}{\tau^2} + \frac{n_j}{\sigma^2})^{-1}$
where $n_j = \#\{i \in \mathcal{L}; z_i = j\}$ and $\bar{y_j} = \left(\sum_{i \in \mathcal{L}; z_i = j} y_i\right) /\,n_j$.

\subsubsection{Full Conditional posterior distribution of $\sigma^2$}

\begin{eqnarray} \label{eq_cond_sigma}
\lefteqn{P(\sigma^2 \mid \mathbf{y}, \boldsymbol{\mu}, \mathbf{z}, \boldsymbol{\theta}, \mathcal{R}) = P(\sigma^2 \mid \mathbf{y}, \boldsymbol{\mu}, \mathbf{z})} \\
&&\propto \left(\frac{1}{\sigma^2}\right)^{1+a} \exp\left\{ - \frac{b}{\sigma^2}\right\} \prod_{i \in \mathcal{L}} \frac{1}{\sqrt{2 \pi \sigma^2}} \exp\left\{-\frac{(y_i - \mu_{z_i})^2}{2 \sigma^2}\right\} \nonumber  \\
&& \propto \left(\frac{1}{\sigma^2}\right)^{1+a+ |\mathcal{L}|/2} \exp\left\{- \frac{1}{\sigma^2} \left(b+ \frac{ \sum_{i \in \mathcal{L}} (y_i - \mu_{z_i})^2}{2}\right) \right\}. \nonumber 
\end{eqnarray}

Therefore, it follows an inverse gamma distribution with parameters $a+ |\mathcal{L}|/2$ and $b+  (1/2)\sum_{i \in \mathcal{L}} (y_i - \mu_{z_i})^2$.

\subsubsection{Full Conditional posterior distribution of $Z_i$}

\begin{equation} \label{eq_cond_z_i}
\begin{aligned}
P(z_i \mid \mathbf{y}, \boldsymbol{\mu}, \sigma^2, \mathbf{z}_{-i}, \boldsymbol{\theta}, \mathcal{R}) &\propto  \frac{\exp \left( \sum\limits_{\br \in \cR} \theta_{z_\bi, z_{\bi + \br}, \br} + \theta_{z_{\bi - \br}, z_i, \br} \right)
 \exp\left(-\frac{(y_i - \mu_{z_i})^2}{2 \sigma^2}\right) }{
    \sum\limits_{a \in \cZ}
    \exp \left( \sum\limits_{\br \in \cR} \theta_{a, z_{\bi + \br}, \br} + \theta_{z_{\bi - \br}, a, \br} \right)
} 
\end{aligned}
\end{equation}

\subsubsection{Full Conditional posterior distribution of $(\btheta, \cR)$}

To address the computational challenges posed by the intractable
normalizing constant of the MRF distribution for $\bz$, we adopt a
pseudolikelihood approximation. Specifically, the likelihood function
$f$ in \eqref{eq:mrf_likelihood} is replaced by the pseudolikelihood
$\tilde{f}$ defined in \eqref{eq:mrf_pseudolik}. This approach is
widely used in MRF models because it eliminates the need to evaluate
the normalizing constant while retaining many of the desirable
optimization properties of the full likelihood.

When incorporated into the Bayesian framework described in the previous
section, the pseudolikelihood induces an alternative target
distribution, referred to as the \emph{pseudoposterior distribution}, as follows
\begin{equation} \label{eq:pseudoposterior}
\begin{aligned}
P(\cR, \btheta  \mid \mathbf{y}, \boldsymbol{\mu}, \sigma^2, \mathbf{z}) &\propto \frac{q(\cR) \phi(\btheta|\cR) \tilde{f}(\bz|\cR, \btheta))
  }{
\sum_{\cR' \in \mathcal{M}} q(\cR') \int_{\mathbb{R}^{d|\cR'|}} \phi(\btheta'|\cR') \tilde{f}(\bz|\cR', \btheta')d\btheta'}.
\end{aligned}
\end{equation}

This distribution is obtained by combining $\tilde{f}$ with the prior
specification of the model parameters and serves as the basis for our
inferential procedures. However, it is important to emphasize that, while
pseudolikelihood may provide a reasonable approximation from an optimization perspective, the two functions can
differ substantially in shape under strong data dependence. Consequently, replacing the likelihood by
the pseudolikelihood alters the interpretation of the resulting
Bayesian quantities. In particular, inference based on the
pseudoposterior should not be viewed as exact Bayesian inference but  alternative measures useful for
assessing uncertainty and comparing competing models. For example, integrating the pseudoposterior distribution over
$\btheta$ produces a measure of relative plausibility across candidate structures $\mathcal{R}$, rather than their exact marginal posterior probabilities.


\subsection{A Reversible Jump Proposal Kernel} \label{sec:RJMCMC}
Constructing a proposal kernel for the Metropolis-Hastings algorithm that can efficiently
move through both the model space $\mathcal{M}$ and the space of interaction coefficients within a model
is not a simple task. In general, RJMCMC  strategy consists of composing a proposal kernel which is a mixture of simpler kernels,
some proposing within-model moves that only changes parameter values and others proposing reversible jumps between models
that have good analytical or computational properties. In this work, we construct a customized proposal kernel, consisting of a mixture of 4 types,  inspired on properties and examples from \citet{brooks2003efficient}
with additional features that are characteristics of the neighborhood selection for MRFs.

\paragraph*{Within-model random walk.} 
The first and simplest move consists of adding a random walk term to the current value of the parameter vector.
Given a current pair $(\cR, \btheta)$, we keep the same neighborhood, $\cR' = \cR$, and propose
a new vector of interaction coefficients $\btheta'$ by adding a Gaussian noise term with matching dimension,
where each coordinate is independent and identically distributed with mean $0$ and variance $\sigma^2_w$,
where $\sigma^2_w$ is a tuning parameter of the algorithm. The transition kernel density for this move is given by
\begin{equation*}
  \kappa_w(\btheta', \cR' | \btheta, \cR) =
  \frac{1}{\left(2\pi\sigma^2_w\right)^{d|\cR|/2}}
  \exp\left(-\frac{1}{2\sigma^2_w} \sum_{\br \in \cR} (\btheta_\br - \btheta_\br')^\top(\btheta_r - \btheta_r') \right)
  \mathds{1}(\cR' = \cR),
\end{equation*}
what makes this proposal density not only reversible, but also symmetrical, i.e.,
$\kappa_w(\btheta', \cR' | \btheta, \cR) = \kappa_w(\btheta, \cR | \btheta', \cR')$.
Since proposed states keep the same interaction structure, we also have $q(\cR) = q(\cR')$, so the terms corresponding
to the neighborhood interaction structure are also cancelled in acceptance ratios.

While this move does not contribute to jumping between RPSs, its goal is to add small incremental changes (small values of $\sigma^2_w$) in the
parameters value so that the chain gradually moves towards higher pseudoposterior density regions within a model. 

\paragraph*{Birth and Death.}
We propose a jump move from a state $(\cR, \btheta)$ to a state $(\cR', \btheta')$, $\cR' \neq \cR$,
by either including a position from $\Rmax$ that is not already in $\cR$ (Birth), or by removing one of the positions in $\cR$ (Death).

Define the RPS comparison operator $\stackrel{\br}{\prec}$ as 
$$\cR\stackrel{\br}{\prec}\cR' \iff  \cR \subset \cR', \br \not\in\cR \text{ and } \cR \cup \{\br \} = \cR',$$
which means that $\cR'$ can be obtained by adding the position $\br$ to $\cR$.
Given a current RPS $\cR$, we randomly select a position $\br^*$ from $\Rmax$ with uniform probabilities
$\frac{1}{|\Rmax|}$. Then either a birth or death move is proposed depending on the selected $\br^*$.

\begin{itemize} 
  \item If $\br^* \in \cR$, the proposed RPS $\cR'$ is such that $\cR' \stackrel{\br^*}{\prec} \cR$,
i.e., $\br^*$ is removed from $\cR$. For the associated interaction coefficients $\btheta'$ to be proposed with $\cR'$, 
all the values are kept the same $\btheta'_\br = \btheta_\br$ for $\br \in \cR'$.
  \item If  $\br^* \not\in \cR$, $\cR'$ is proposed by including $\br^*$, i.e., $\cR \stackrel{\br^*}{\prec}\cR'$.
    For the proposed parameter $\btheta'$, we keep the values of the previous $\btheta'_\br = \btheta_\br$ for
    the previously included positions $\br \in \cR$ and sample a new vector of i.i.d. Gaussian variables with
    mean $0$ and variance $\sigma^2_{\text{bd}}$ to assign to $\btheta'_{\br^*}$.
\end{itemize}
Note that, by this definition, transitions between states with two different RPSs $\cR$ and $\cR'$ are allowed
if, and only if, $|\cR \triangledown \cR'| = 1$, where $\triangledown$ denotes the symmetrical difference
operator for two sets.
Therefore, the proposal kernel density for a birth/death jump move is given by
\begin{equation*}
  \kappa_{\text{bd}}(\btheta', \cR' | \btheta, \cR) = 
  \begin{cases}
  \frac{1}{|\Rmax|}
  \frac{\exp\left(-\frac{1}{2\sigma^2_\text{bd}} {\btheta'}_{\br^*}^\top {\btheta'}_{\br^*} \right)}{\left(2\pi\sigma^2_{\text{bd}}\right)^{d/2}} 
    \prod_{\br \in \cR}\mathds{1}(\btheta'_\br = \btheta_\br) &, \text{ if }\cR \stackrel{\br^*}{\prec} \cR', \\
    \frac{1}{|\Rmax|} \prod_{\br \in \cR'}\mathds{1}(\btheta'_\br = \btheta_\br) &, \text{ if }\cR'  \stackrel{\br^*}{\prec} \cR, \\
    0 &, \text{ if } |\cR \triangledown \cR'| \neq 1.
  \end{cases}
\end{equation*}

\begin{figure}[ht!]
  \centering
  \includegraphics[width = 0.8\linewidth]{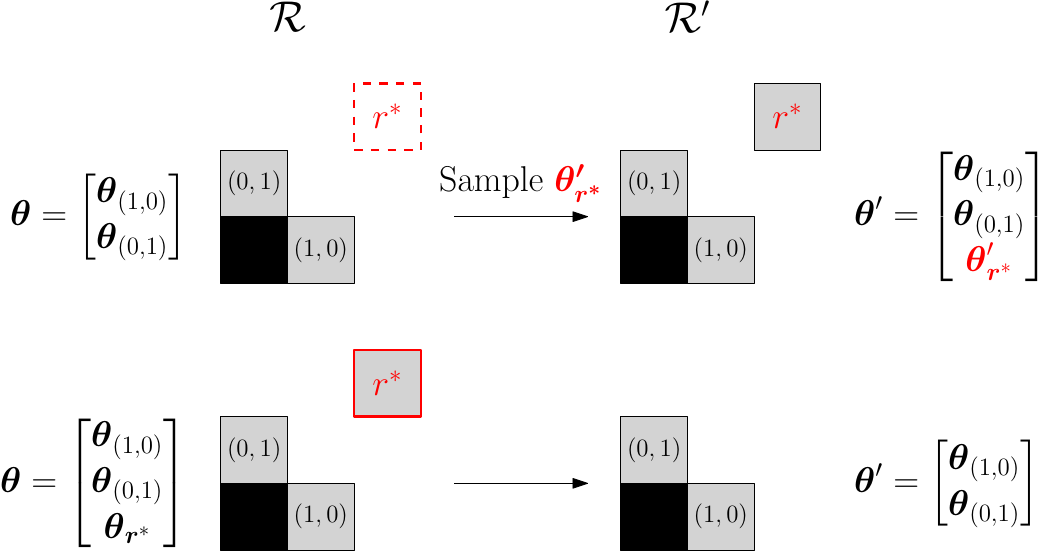}
  \caption{Illustration of a Birth/Death Jump proposal when sampling $\br^* \not\in \cR$ (top) and $\br^* \in \cR$ (bottom).}\label{fig:illustration_jump}
\end{figure}

\autoref{fig:illustration_jump} illustrates how new states $(\cR', \btheta')$ are proposed from a current $(\cR, \btheta)$
when a randomly selected position $\br^*$ is included or not included in $\cR$.
It is straightforward to conclude from the example that this type of jump can be reversed by selecting the same position $\br^*$ and,
the case of adding a new position, sampling the appropriate $\btheta_{\br^*}'$, therefore,
$\kappa_\text{bd}(\btheta', \cR' | \btheta, \cR) > 0$ if, and only if, $\kappa_\text{bd}(\btheta, \cR | \btheta', \cR') > 0$.

\begin{figure}[ht!]
  \centering
  \includegraphics[width = 0.8\linewidth]{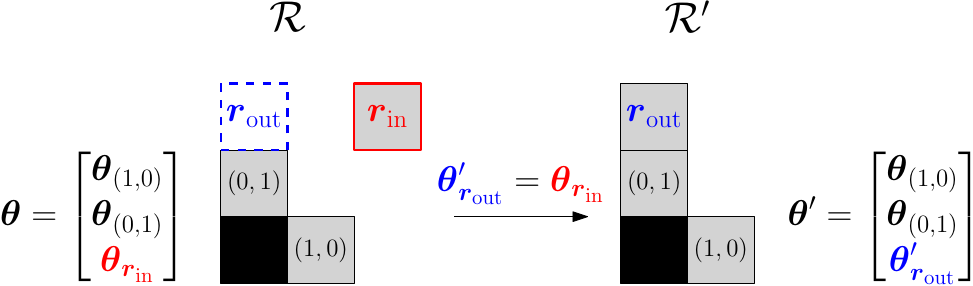}
  \caption{Illustration of a Position swap move proposal.}\label{fig:illustration_swap}
\end{figure}

\paragraph*{Position swap}
As second type of move constructed for proposing jumps between states with different RPS is to swap one of positions in the current RPS,
$\br_{\text{in}} \in \cR$, for another one $\br_{\text{out}} \in \Rmax \setminus \cR$,
while keeping all of the interaction coefficients the same, including the subvector associated with the swapped position,
$\btheta'_{\br_\text{out}} = \btheta_{\br_{\text{in}}}$.
\autoref{fig:illustration_swap} illustrates an example of \textbf{Position swap move} proposal.

The positions $\br_{\text{in}}$ and $\br_{\text{out}}$, for the swap move, are chosen independently and uniformly distributed on $\cR$ and $\Rmax \setminus \cR$, respectively.
Therefore, for any states $(\btheta, \cR)$ and $(\btheta', \cR')$ such that $|\cR| = |\cR'|$ and 
$|\cR \triangledown \cR'| = 2$, differing only in the presence of relative positions $\br_\text{in} \in \cR$ and 
$\br_\text{out} \in \cR'$, the proposal density for the swap move is given by
\begin{equation}\label{eq:move_sw}
  \kappa_{\text{sw}}(\btheta', \cR' | \btheta, \cR) =
    \frac{1}{|\cR| |\Rmax \setminus \cR|}
    \prod\limits_{\br \in \cR \cap \cR'} \mathds{1}(\btheta_\br = \btheta_\br')
    \mathds{1}(\btheta_{\br_\text{in}} = \btheta'_{\br_{\text{out}}}).
\end{equation}
Note that \eqref{eq:move_sw} is a symmetrical kernel since the inverse operation is proposed by selecting the same pair of positions
$\br_\text{in}$ and $\br_\text{out}$ reversed, the RPS prior probability only depends on the size of the current RPS, $|\cR|$, and we have
the condition that $|\cR| = |\cR'|$ for every pair of states with positive proposal probability.

The rationale behind this move is that, due to the spatial dependence intrinsic to lattice-based indexing of the MRF model considered,
the counts of pairwise configurations in some relative positions may present high correlation,
especially when those relative positions are close (e.g. $\br$ and $\br + (1,0)$) or a multiple one from another (e.g. $\br$ and $2\br$).

Note that while the same jumps proposed by swap moves could be achieved by a series of birth and death moves, one of those steps would be to
exclude a relative position, $\br_\text{in}$, with associated interaction coefficient, $\btheta_{\br_\text{in}}$, possibly far from the zero
vector, what would cause the acceptance of such move to be highly unlikely. Thus, swap moves is a proposal step that avoids the
algorithm getting stuck at a local (with respect to RPSs) maxima, by adding direct connections to states with different RPSs that
may have similar pseudoposterior values, taking into account very specific characteristics of the model.

\begin{figure}[ht!]
  \centering
  \includegraphics[width = 0.9\linewidth]{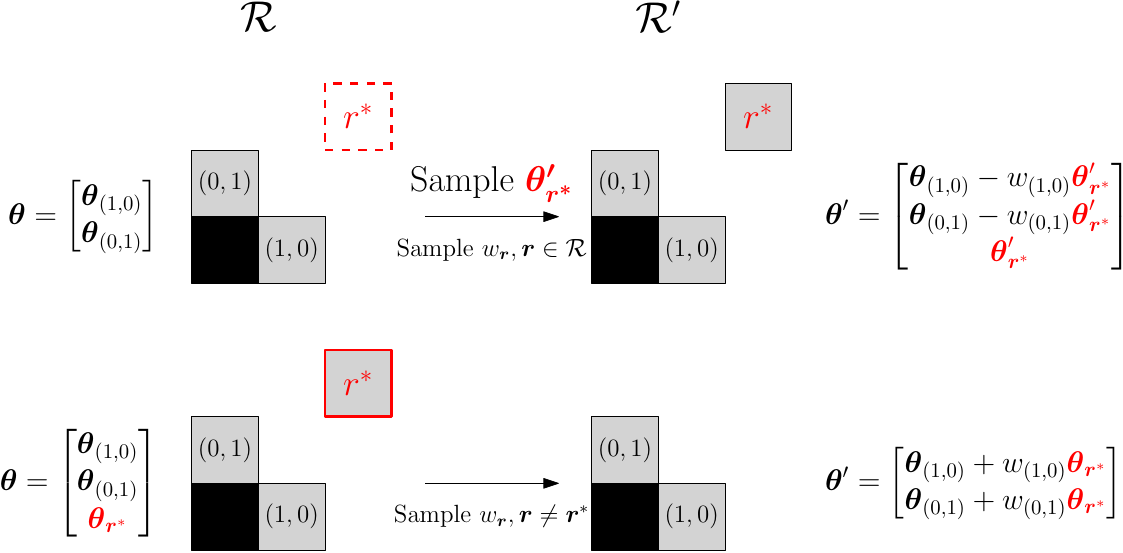}
  \caption{Illustration of Split (top) and Merge (bottom) moves.}\label{fig:illustration_merge}
\end{figure}

\paragraph*{Split and Merge.}
While the Position Swap Move is designed to replace an active relative position with an inactive candidate possessing a similar pseudoposterior value, local maxima can also arise when a model simultaneously includes multiple relative positions with highly correlated sufficient statistics. High correlation between these sufficient statistics can render the interaction weights $\boldsymbol{\theta}_{\mathbf{r}}$ unstable, as large parameter shifts in one direction can be offset by compensating shifts in another without substantially altering the model fit. To address this instability and enable transitions across candidate relative position structures (RPSs), we introduce the Split and Merge moves as a pair of reversible operations that adjust both model dimensionality and parameter values, see Figure \ref{fig:illustration_merge}. By redistributing interaction coefficients $\boldsymbol{\theta}_{\mathbf{r}}$, these moves facilitate proposals between smaller and larger RPSs of similar likelihood. Specifically, the Merge move removes redundant relative positions from the current state by reallocating their interaction weights to the remaining positions, thereby maintaining pseudolikelihood levels while keeping the pseudolikelihood values
on similar levels. The key premise on this pair of moves is that, for a pair of states $(\cR, \btheta)$ and $(\cR', \btheta')$,
we have 
$$\sum_{\br \in \cR} \btheta_\br = \sum_{\br \in \cR'} \btheta'_{\br'}.$$

Given a current state $(\cR, \btheta)$, the process of proposing a state $(\cR', \btheta')$ with a \textbf{Split} move
is composed by the steps of 
\begin{enumerate}
  \item Sample a new position $\br^*$ to be included from $\Rmax \setminus \cR$ with uniform probability.
  \item Generate a new interaction coefficient vector $\btheta'_{\br^*}$ from a $d$-dimensional independent
    Gaussian distribution with variances $\sigma^2_s$, where the split variance, $\sigma^2_s$,
    is a tuning parameter of the algorithm.
  \item Generate a vector of weights $\boldsymbol{w} = (w_\br)_{\br \in \cR}$ from a
    symmetric Dirichlet distribution with all parameters equal to $\nu$,
    where $\nu$ is another tuning parameter of the algorithm.
  \item Propose $\cR' = \cR \cup \br^*$ and $\btheta'$ such that $\btheta'_{\br^*}$ is the generated vector
    and $\btheta'_{\br} = \btheta_{\br} -w_\br \btheta'_{\br^*}$ for every other relative position $\br \in \cR$.
\end{enumerate}
The proposal density for a Split move, $\kappa_{\text{s}}$, is given by
\begin{align*}
  \kappa_{\text{s}}(\btheta', \cR' | \btheta, \cR) = 
  &
  \frac{\prod\limits_{\br^* \in \Rmax \setminus \cR} \mathds{1}(\cR \stackrel{\br^*}{\prec} \cR')}{|\Rmax \setminus \cR|}
\frac{
  \exp\left( \frac{-({\btheta'_{\br^*}}^\top\btheta'_{\br^*})}{2\sigma^2_s}\right)
}{
  \left(2\pi\sigma^2_s\right)^{d/2}  
}\\
&\frac{\Gamma\left(|\cR|\nu\right)}{\prod\limits_{\br \in \cR}\left(\Gamma(\nu)\right)}
\prod_{\br \in \cR} \left[\left( \frac{\theta_{0,1,\br} - \theta'_{0,1,\br}}{\theta'_{0,1,\br^*}}\right)^{\nu-1} \times 
\frac{\mathds{1} \left( 0 < \frac{\theta_{0,1,\br} - \theta'_{0,1,\br}}{\theta'_{0,1,\br^*}} < 1 \right)}{\theta'_{0,1,\br^*}}\right]\\
& \mathds{1}\left( \sum_{\br \in \cR} \left(\btheta_\br - \btheta'_\br\right) = \btheta'_{\br^*} \right)
\prod_{a,b \in \cZ^2} \mathds{1}\left(\frac{\theta_{a,b,\br} - \theta'_{a,b,\br}}{\theta'_{a,b,\br^*}} = \frac{\theta_{0,1,\br} - \theta'_{0,1,\br}}{\theta'_{0,1,\br^*}}\right).
\end{align*}

The \textbf{Merge} proposal move is the reverse of the Split move and can be described by the following sequence of steps
\begin{enumerate}
  \item Sample a state $\br^*$ from $\cR$, which interaction coefficient vector will be merged into the others,
    with uniform probabilities.
  \item Generate a vector of weights $\boldsymbol{w} = (w_\br)_{\br \in (\cR \setminus \br^*)}$ with Dirichlet distribution with all parameters equal to $\nu$.
  \item Propose $\cR' = \cR \setminus \br^*$ and $\btheta'$ such that $\btheta'_\br = \btheta_\br + w_{\br}\btheta_{\br^*}$ 
    for each $\br \in \cR'$.
\end{enumerate}
The proposal density for a Merge move is
\begin{align*}
  \kappa_{\text{m}}(\btheta', \cR' | \btheta, \cR) = 
  &
  \frac{\prod\limits_{\br^* \in \cR}\mathds{1}(\cR' \stackrel{\br^*}{\prec} \cR)}{|\cR|}
  \\
  &
  \frac{\Gamma\left(|\cR'|\nu\right)}{\prod\limits_{\br \in \cR'}\left(\Gamma(\nu)\right)}
  \prod_{\br \in \cR'} \left[ \left(\frac{\theta'_{0,1,\br} - \theta_{0,1,\br}}{\theta_{0,1,\br^*}}\right)^{\nu - 1}
  \times \frac{\mathds{1}\left(0 <\frac{\theta'_{0,1,\br} - \theta_{0,1,\br}}{\theta_{0,1,\br^*}} < 1\right)}{\theta_{0,1,\br^*}} \right]
  \\
  & 
  \mathds{1}\left(\sum_{\br \in \cR'} \btheta_\br' - \btheta_\br = \btheta_{\br^*}\right)
  \prod_{a,b \in \cZ^2} \mathds{1}\left(\frac{\theta_{a,b,\br} - \theta'_{a,b,\br}}{\theta'_{a,b,\br^*}} = \frac{\theta_{0,1,\br} - \theta'_{0,1,\br}}{\theta'_{0,1,\br^*}}\right).
  \end{align*}

It is easy to see that an accepted Split move is reversed by a Merge move
if the sampled relative position $\br^*$ and the generated vector of weights $\boldsymbol{w}$ is the same
for both moves.
This fact also produces an analytical simplification in the ratio of proposal densities,
required for computing the acceptance ratio in the Metropolis-Hastings algorithm, for these moves as
\begin{equation}\label{eq:split_merge_ratio}
  \frac{\kappa_{\text{m}}(\btheta, \cR | \btheta', \cR')}{\kappa_{\text{s}}(\btheta', \cR' | \btheta, \cR)} =
  \frac{|\Rmax \setminus \cR|}{|\cR'|} 
  \frac{
  \exp\left( \frac{-({\btheta'_{\br^*}}^\top\btheta'_{\br^*})}{2\sigma^2_s}\right)
}{
  \left(2\pi\sigma^2_s\right)^{d/2}  
}
  ,
\end{equation}
for any pair of states $(\cR, \btheta)$ and $(\cR', \btheta')$ such that $\kappa_{\text{s}}(\btheta', \cR' | \btheta, \cR) > 0$.
The ratio of proposal densities for the reversed transitions can be achieved by applying the inverse of \eqref{eq:split_merge_ratio}.

\paragraph*{Mixture proposal density.}
Each move described previously has a different goal in terms of how a region (in terms of both RPS and interaction coefficients)
of much higher posterior density could be proposed with some probability improving the rate of convergence of the
Metropolis-Hastings algorithm to a region corresponding to a global maximum.
In order to assemble the five groups of moves into a single transition kernel, we define a proposal density $\kappa$ that is composed
by a mixture of the of five described densities
\begin{equation}
  \kappa(\btheta', \cR' | \btheta, \cR) = \sum_{\Psi \in \{\text{w,bd,sw,m,s}\}} p_{\Psi}(\cR)\kappa_{\Psi}(\btheta', \cR' | \btheta, \cR),
\end{equation}
where $p_\Psi(\cR)$ corresponds to the probability of selecting a move $\Psi$ when the current state has the RPS component as $\cR$
and $\sum_{\Psi} p_{\Psi}(\cR) = 1$. Having the mixture probabilities depend on the current RPS is required to avoid undefined behaviors
such as proposing a random walk move $\kappa_\text{w}$ when we have an empty RPS, $\cR = \emptyset$.

Following \citet{green1995reversible} and \citet{brooks2003efficient}, we can describe the acceptance of the Metropolis-Hastings
(Reversible Jump) algorithm when using a mixture proposal density by using the ratio of the sampled move $\Psi$ at each step,
\begin{equation}\label{eq:acceptance}
  \mathcal{A}_{\Psi}(\btheta', \cR' | \btheta, \cR) = 
  \frac{q(\cR')}{q(\cR)}
  \frac{\pi(\btheta'|\cR')}{\pi(\btheta|\cR)}
  \frac{\tilde{f}(\btheta', \cR')}{\tilde{f}(\btheta, \cR)}
  \frac{p_{\Psi'}(\cR)}{p_{\Psi}(\cR)}
  \frac{\kappa_{\Psi'}(\btheta, \cR | \btheta', \cR')}{\kappa_\Psi(\btheta', \cR' | \btheta, \cR)},
\end{equation}
where $\Psi'$ is the inverse move of the move $\Psi$, i.e., $\Psi' = \Psi$ if $\Psi = \text{w, bd or sw}$ and
swapped for $\Psi = \text{s or m}$. 
In practice, computations involving $\mathcal{A}_{\Psi}(\cdot, \cdot | \cdot, \cdot)$ can be performed
in logarithmic scale for both analytical and numerical simplicity.


\section{Simulation Studies} \label{sec:simulation}

We conducted a simulation study to evaluate the proposed reversible jump procedure in the setting where the underlying MRF is latent and only noisy observations of the latent field are available. The simulation study has two main objectives. First, we investigate the recovery of the latent spatial structure and its associated relative-position set (RPS) from noisy observations. Second, we examine the sensitivity of the inferred interaction structure to the complexity penalty used by the reversible jump procedure.

\subsection{Simulation design}

All simulations were performed on a regular two-dimensional lattice of size $150\times150$. For each simulation scenario, a latent field $ \bZ=(Z_s:s\in\mathcal S)$ with state space $\mathcal Z=\{0,1,2\}$  was first generated according to a prespecified MRF model. The observed field $\bY$ was then generated conditionally on $\bZ$ according to

$$
Y_s=Z_s+\varepsilon_s,
\qquad
\varepsilon_s\overset{\mathrm{iid}}{\sim}N(0,1),
\quad s\in\mathcal S.
$$

Therefore, the spatial dependence is generated at the latent level, whereas the observations available for inference are noisy continuous measurements of the latent states. In particular, the realization of $\bZ$ is not provided to the inference procedure. It is retained only as the known ground truth for evaluating the simulation.

\subsection{Data-generating models}

We considered two data-generating scenarios with different interaction structures. In the first scenario, the latent MRF contains only first-order interactions, with relative-position set $
\mathcal R_1=\{(1,0),(0,1)\},$  and interaction parameters $
\btheta_1=(-1.5,-1.5)$. In the second scenario, the latent MRF contains the same first-order interactions together with an additional interaction at relative position $(3,3)$. That is, $\mathcal R_2=\{(1,0),(0,1),(3,3)\}$, with interaction parameters 
$\btheta_2=(-1.7,-1.7,0.4)$. 

For each scenario, the latent field was generated using the corresponding RPS and interaction parameters. The observed field was subsequently obtained by adding independent standard Gaussian noise at each lattice site.

The two scenarios were chosen to provide a simple baseline and an extended interaction structure. The first scenario considers a model with only the nearest-neighbor interactions, whereas the second introduces an additional, more distant relative position. The comparison therefore allows us to investigate the recovery of the underlying interaction structure when the spatial field generating the observations is itself unobserved.

\subsection{Simulated data}

\begin{figure}[ht]
    \centering
    \includegraphics[width=0.8\textwidth]{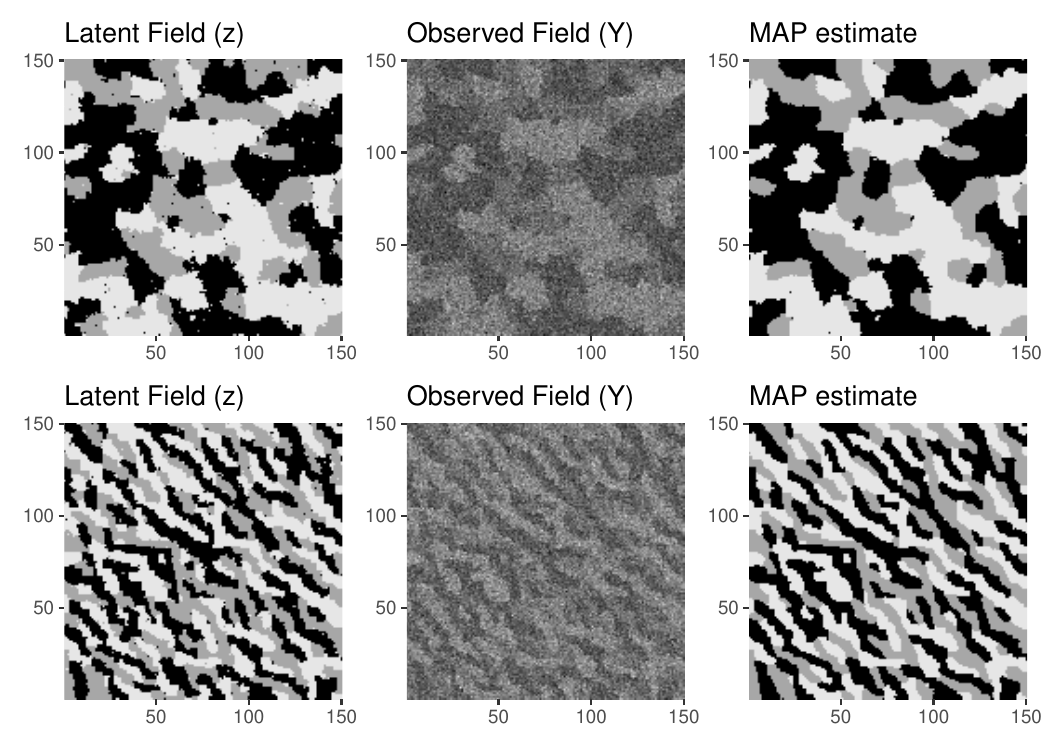}
    \caption{Latent fields, noisy observations, and MAP estimates for the two simulation scenarios. The first row corresponds to Simulation 1 with RPS $\mathcal R_1$, and the second row corresponds to Simulation 2 with RPS $\mathcal R_2$. The first column shows the latent field $\bZ$, the second column shows the observed field $\bY$, and the third column shows the MAP estimate of $\bZ$.}\label{fig:simulation-fields}
\end{figure}

Figure~\ref{fig:simulation-fields} illustrates the data-generating and inferential setup for the two simulation scenarios. For each scenario, the figure displays the latent field $\bZ$, the observed noisy field $\bY$, and the MAP estimate of the latent field obtained from the fitted model.

This figure provides a visual representation of the inferential problem considered here: the spatial dependence is present in the latent field, but the model must infer both the latent configuration and the interaction structure from its noisy observations.

\subsection{Inference procedure}

For both simulation scenarios, inference was performed using the proposed Bayesian reversible jump procedure. The maximal interaction structures were specified as a RPS containing all relative positions within a $3\times3$ neighborhood of the origin,
and the sampler was initialized using the nearest-neighbor structure.
In all runs, we performed 5000 steps of the within-model random walk move under the nearest-neighbor RPS as a warm-up before starting the reversible jump procedure. This initialization was intended to provide a reasonable starting point for the sampler, allowing it to explore the posterior distribution more effectively.

We also considered a restriction that $\theta_{r,a,b} = 0$ whenever $a = b$ and $\theta_{r,a,b} = \theta_r$ when $a \neq b$.
This restriction reduces the number of parameters to be estimated and reflects the assumption that interactions are only present between different states.
The proposal standard deviation for the kernel parameters was set to $0.01$, and the prior standard deviation for the $\theta_r$ parameters was set to $10$.
For the RPS prior distribution, we used the prior $q$  with hyperparameters $\alpha = 1$ and $\beta = \log(150 \times 150)$. This prior specification introduces a complexity penalty, related to the grid size ($|\cS|= 150 \times 150$), that discourages overly complex interaction structures, while still allowing for flexibility in the inferred RPS. A value of $\alpha = 1$ makes this penalty similar to what would be obtained from a BIC-type criterion, which is commonly used in model selection contexts.

For each scenario, the sampler was run for 50,000 iterations, with the first 5,000 warm-up iterations discarded. Posterior inference was based on the remaining iterations. Importantly, all quantities required by the inference procedure were computed from the observed field $\bY$. The latent field $\bZ$ was not supplied to the sampler.

\subsection{Computational implementation}

All computations were carried out in R, using the \textbf{mrf2dbayes} package
(\url{https://github.com/Freguglia/mrf2dbayes}), which implements the proposed
Bayesian inference procedure.
The package builds on the data structures provided by the \textbf{mrf2d} package
\citep{freguglia2020inference} and extends it with Bayesian analysis elements,
including the pseudolikelihood approximation and the Metropolis-Hastings and
reversible jump algorithms described in this work.

\subsection{Evaluation}

The simulation study considers two complementary aspects of inference.
First, we examine reconstruction of the latent field. The posterior probabilities of the latent states are obtained for each lattice site, and these probabilities are summarized through the corresponding MAP configuration, $\widehat{\bZ}_{\mathrm{MAP}}$.
The estimated latent field can then be compared with the known realization used to generate the observed data. Second, we examine recovery of the interaction structure. The reversible jump sampler explores models corresponding to different RPSs, allowing posterior information about the inclusion of candidate relative positions to be obtained. This posterior information is compared with the RPS used in the data-generating model.

\begin{figure}[ht]
    \centering
    \includegraphics[width = 0.6\linewidth]{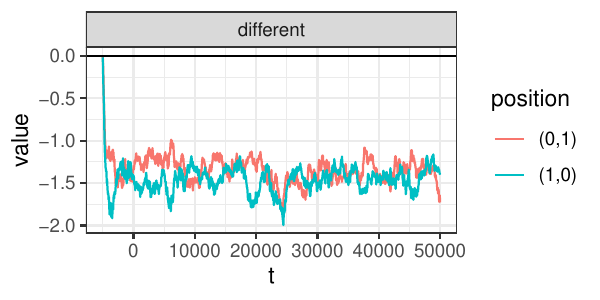}
    \includegraphics[width = 0.3\linewidth]{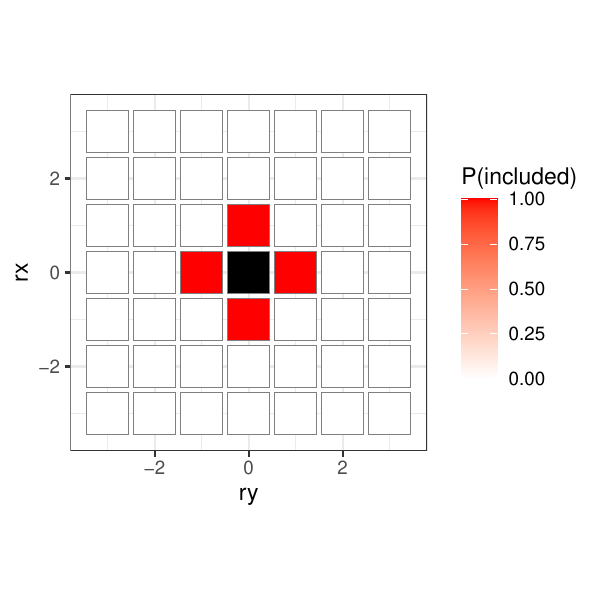}
    \caption{Posterior trace of $\btheta$ and distribution for the interaction positions in $\cR$ in Simulation 1. The left panel shows the trace plots for each parameter, while the right panel shows the corresponding posterior inclusion probability of each position in the maximal RPS.}\label{fig:sim1_theta}
\end{figure}

\begin{figure}[ht]
    \centering
    \includegraphics[width = 0.6\linewidth]{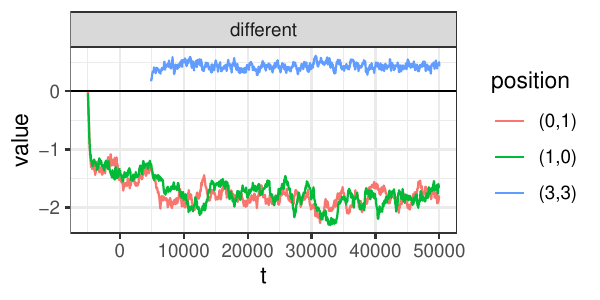}
    \includegraphics[width = 0.3\linewidth]{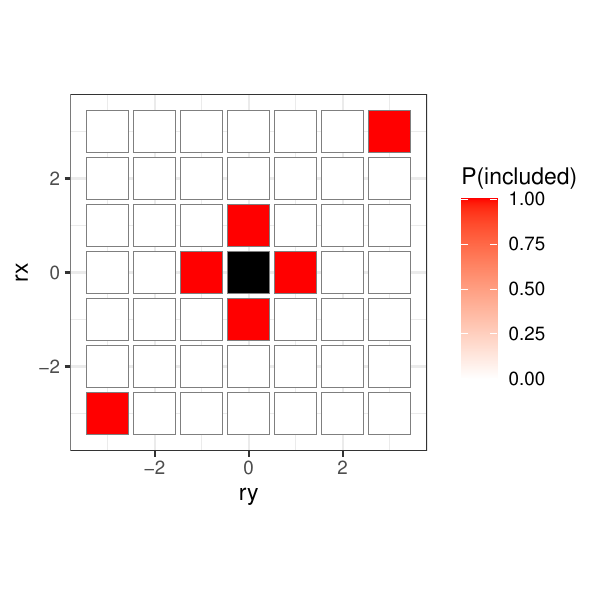}
    \caption{Posterior trace of $\btheta$ and distribution for the interaction positions in $\cR$ in Simulation 2. The left panel shows the trace plots for each parameter, while the right panel shows the corresponding posterior inclusion probability of each position in the maximal RPS.}\label{fig:sim2_theta}
\end{figure}

For the main simulation experiment, Figures \ref{fig:sim1_theta} and  \ref{fig:sim2_theta} present posterior traces and distributions of the model parameters, the posterior behavior of the candidate relative positions, the posterior probabilities of the latent states, and the resulting MAP latent-field configuration.

\subsection{Sensitivity to the complexity penalty}

We conducted a separate simulation experiment to investigate the sensitivity of the interaction-structure inference to the complexity penalty induced by the prior hyperparameter $\alpha$.
For this experiment, we fixed the data-generating mechanism to the second scenario,
$\mathcal R_2=\{(1,0),(0,1),(3,3)\}$,
with the same interaction parameters.


We considered six values for the hyperparameter $\alpha$ in the prior distribution of the RPS:
$
\alpha\in\{0.1,0.5,1,2,5,10\}.
$
For each value of $\alpha$, the complexity penalty induced by the prior was specified as
$\alpha\log(|\mathcal S|)$. Each chain was run for 25,000 iterations, with the first 5,000 iterations discarded as warmup, starting from the same nearest-neighbor configuration as the main simulation. All other aspects of the model and the simulation setup were kept fixed across the different values of $\alpha$.

\begin{figure}[ht]
    \centering
    \includegraphics[width = 0.8\textwidth]{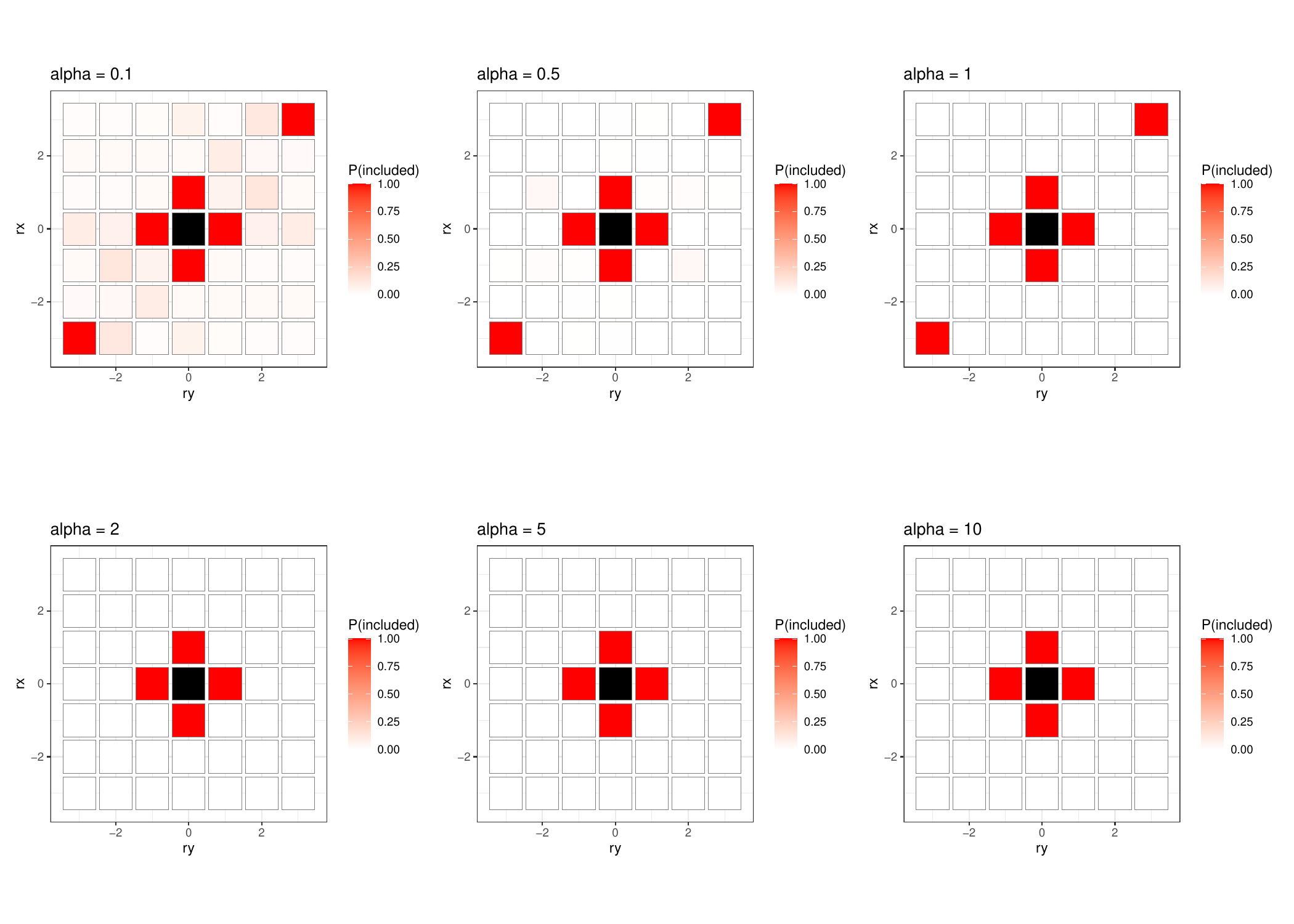}
    \caption{Posterior inclusion probability of each position in $\Rmax$ for the run under each value of $\alpha$ considered.} \label{fig:alpha}
\end{figure}

\begin{figure}[ht]
    \centering
    \includegraphics[width = 0.8\textwidth]{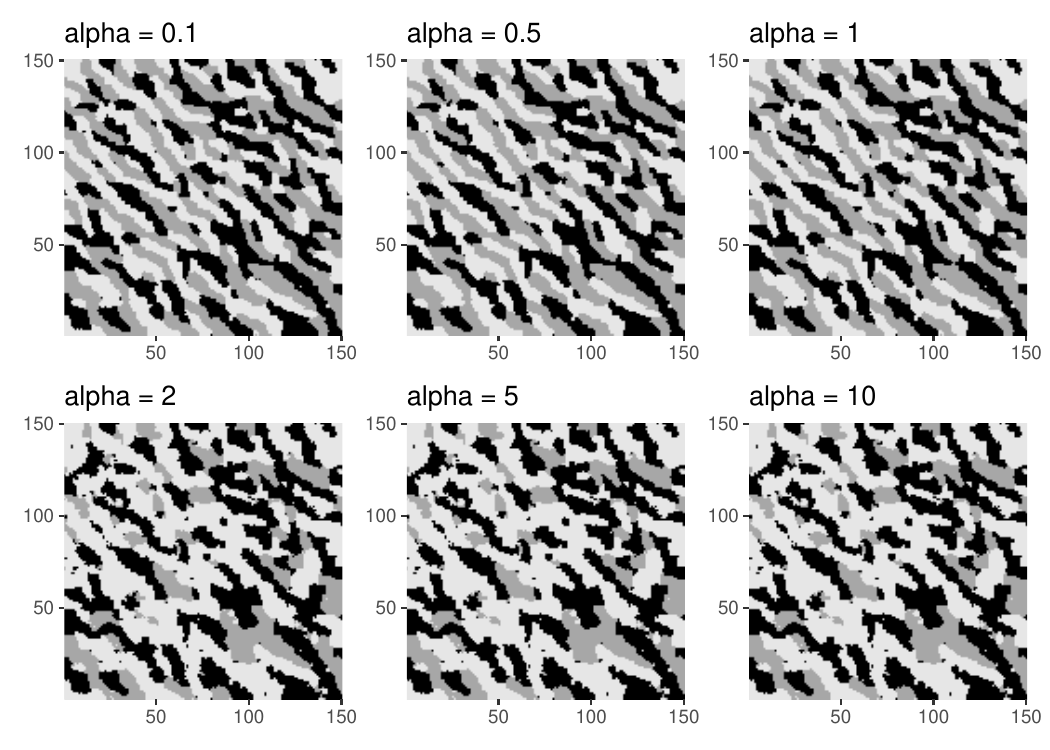}
    \caption{Segmentation of the latent field for each value of $\alpha$ considered.} \label{fig:latent}
\end{figure}

This experiment investigates how the strength of the complexity penalty affects the posterior selection of the interaction structure. Specifically, we examine the posterior support for candidate relative positions and the corresponding maximum a posteriori (MAP) estimates of the latent field across various values of $\alpha$. Figures~\ref{fig:alpha} and \ref{fig:latent} show that small values of $\alpha$ under-penalize model complexity, leading to an overestimated neighborhood. Conversely, large values of $\alpha$ lead to over-penalization, retaining only the immediate nearest neighbors; while this reflects strong local dependence, it results in poor spatial segmentation.

\section{Application to HTP images}\label{sec:application}

To illustrate the proposed model selection methodology in a real-data setting, we apply the proposed procedure to an image acquired in a high-throughput phenotyping (HTP) context using the image from \autoref{fig:planta}. The objective is to investigate the spatial structure identified by the model and to obtain a latent representation of the image through the inferred MRF, that can be interpreted as a segmentation of the image into different classes.
This segmentation allows us to identify regions of the image that represents the characteristics of the image such as the presence of vegetation, soil, and other features.


The original image is a color image with red, green, and blue (RGB) channels. To obtain a scalar field suitable for the proposed MRF framework, we transform the RGB image using the Excess Green Vegetation Index (ExG). Specifically, denoting the red, green, and blue channels by $R$, $G$, and $B$, respectively, we define
$$Y = 2G-R-B.$$

\begin{figure}
\centering
\includegraphics[width=0.5\textwidth]{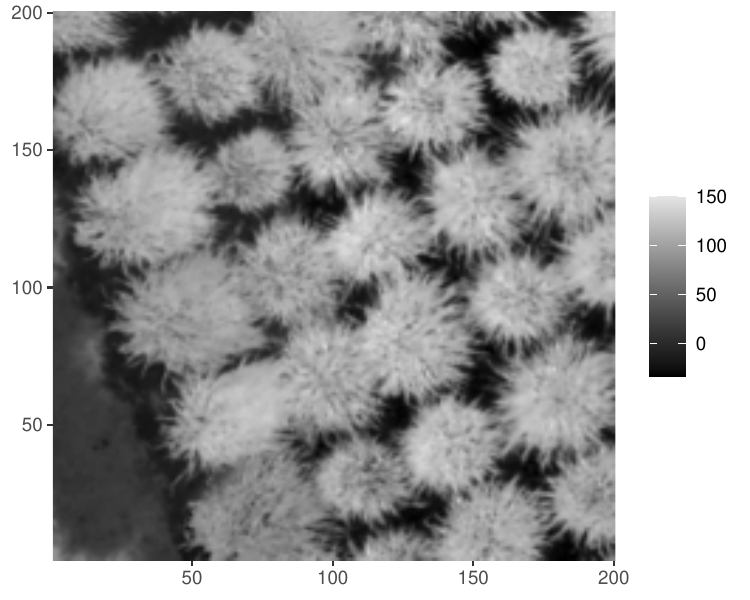}
\caption{Excess Green Vegetation Index (ExG) representation of the
region used in the analysis.}\label{fig:apli_y}
\end{figure}

The resulting field is then restricted to a $200\times200$ region of the original image. Figure~\ref{fig:apli_y} shows the resulting ExG representation used as input to the model.

The resulting ExG field is the observed field $\bY$ used in the subsequent analysis. In contrast to the simulation study, the latent spatial field and its generating interaction structure are unknown in this application.



We consider the maximal interaction structure $\Rmax$, similarly to those in the simulation study, as the structure including all positions within a $3 \times 3$ grid. Among the model specifications considered in the preliminary analysis, we focus on the model with three latent classes and a symmetric interaction family, i.e., we include a restriction on the interaction parameters such that $\theta_{r,a,b} = \theta_{r,b,a}$.
The symmetric interaction family is used to impose the corresponding symmetry constraints on the interaction parameters, reducing the number of free parameters while retaining the spatial dependence represented by the selected relative positions. We use a standard deviation of $0.025$ for the proposal moves involving the interaction parameters, and prior standard deviation $1$. The chain is run for 25,000 iterations, with the first 5,000 iterations discarded as warmup, also restricted to the nearest-neighbor interaction structure during the warmup iterations.


Since the true latent field is unavailable for the real image, model assessment cannot be based on direct comparison with a known generating configuration. Instead, we examine the inferred latent representation and the posterior distribution over interaction structures.

\begin{figure}[ht]
    \centering
    \includegraphics[width=0.5\textwidth]{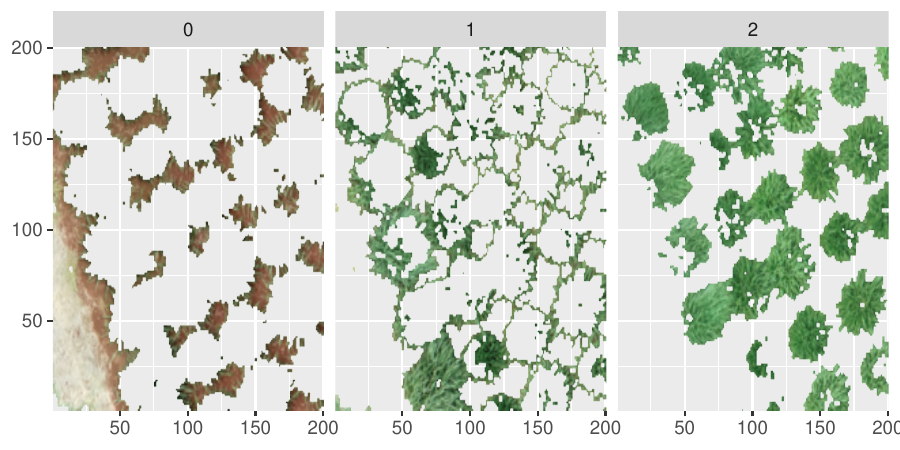}
    \caption{MAP configuration of the latent field for the three-class symmetric model. Pixels of the original (color) image are shown within the corresponding segments.}\label{fig:apli_seg}
\end{figure}

The latent representation is summarized by the MAP configuration,
$\widehat{\bZ}_{\mathrm{MAP}}$,
obtained from the posterior state probabilities at each pixel. This produces a segmentation of the image into the three latent classes.
We visualize the MAP configuration in Figure~\ref{fig:apli_seg}, where the pixels of the original color image are shown within the corresponding segments.

\begin{figure}[ht]
    \centering
    \includegraphics[width=0.4\textwidth]{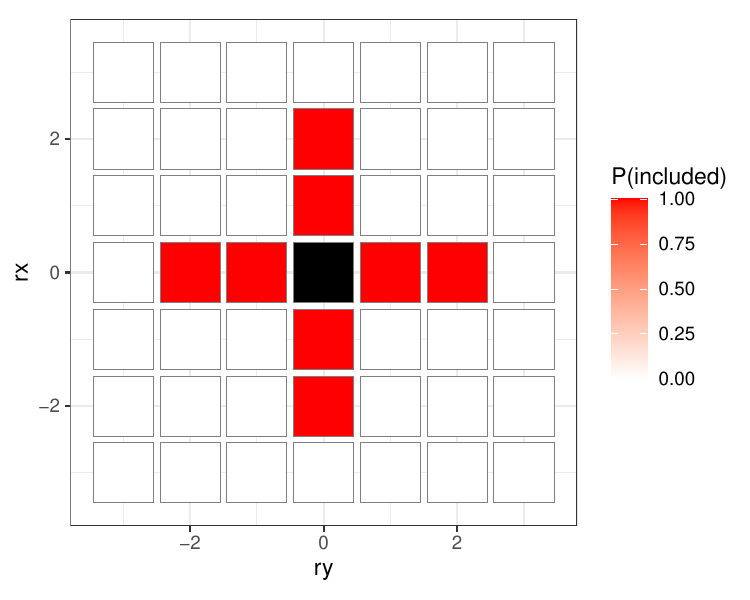}
    \caption{Posterior distribution over candidate relative positions for the three-class symmetric model.}\label{fig:mrfi}
\end{figure}

The inferred interaction structure is summarized through the posterior distribution over candidate relative positions in Figure~\ref{fig:mrfi}.
In particular, the posterior probabilities associated with the candidate interactions indicate which spatial offsets are supported by the observed image. 

\begin{figure}[ht]
    \centering
    \includegraphics[width=0.8\textwidth]{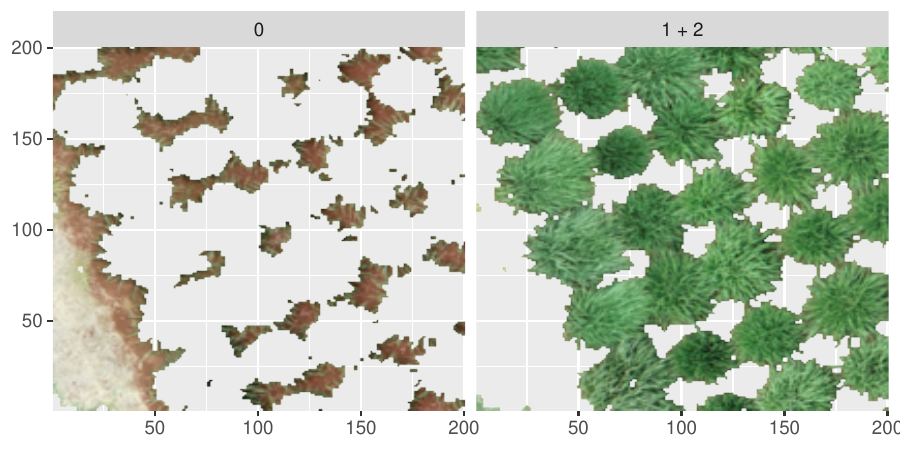}
    \caption{Pixels of the original (color) image within the corresponding segments, joining classes 1 and 2.}\label{fig:apli_seg_merged}
\end{figure}

\section{Conclusion}

In this paper, we presented a Bayesian image segmentation approach using  a Gaussian mixture likelihood driven by a latent Markov Random Field (MRF), introducing a novel mechanism for adaptive spatial structure learning. Unlike traditional MRF-based segmentation architectures that rely on pre-fixed neighborhood configurations, our approach operates within a fully Bayesian paradigm that treats the neighborhood structure itself as an unknown parameter vector. By leveraging Reversible Jump Markov Chain Monte Carlo (RJMCMC) sampling, the framework dynamically transits between neighborhood configurations of varying dimensions and topologies, yielding robust spatial clustering and superior segmentation accuracy without manual tuning of neighborhood dependencies. 

While High-Throughput Phenotyping (HTP) served as our primary motivating application, where accurate plant organ and canopy extraction from complex backgrounds is essential, the framework is generalizable to a wide range of computer vision and spatial modeling problems. For instance, the underlying formulation can be naturally adapted to model surface textures and defect dynamics in industrial materials, such as textile texture modeling, where the interlacing of warp and weft yarns, combined with subtle fabric shifts, generates a texture effect that induces a complex spatial dependency structure in pixel colors (see for example \citep{freguglia2020hidden}).

On the other hand, the modelling can be generalized for bounded and non-Gaussian mixtures. For example, in agricultural and environmental remote sensing contexts, imagery often incorporates continuous vegetation metrics such as the Normalized Difference Vegetation Index (NDVI), which is constrained to the interval $[-1, 1]$. Extending our framework to accommodate bounded or skewed distributions (e.g., Beta or truncated distributions) would improve modeling fidelity for such indices. Methodologically, replacing Gaussian components with non-conjugate mixture distributions requires substituting standard Gibbs updates with Metropolis-Hastings steps for the mixture parameters, introducing a modest computational trade-off in exchange for heightened flexibility. Also, the prior distribution over the space of candidate MRF neighborhoods can be enriched to encode spatial physics or domain-specific geometric constraints. For example, we can use polynomial penalization functions based on the Euclidean or geodesic distance between interacting pixel neighbors, discouraging long-range or overly dense configurations unless strongly supported by the spatial data likelihood.

\if1\anon
\section{Acknowledgments}\label{acknowledgments}

The authors gratefully acknowledge the Fundação de Amparo à Pesquisa do
Estado de São Paulo (FAPESP), grants 2017/25469-2, 2022/04006-2 and 2023/06910-0, CNPq grants 304148/2020-2 and 306496/2024-0, and
the Coordenação de Aperfeiçoamento de Pessoal de Nível Superior -- Brasil
(CAPES), Finance Code 001.
\fi






\bibliography{biblio.bib}
\bibliographystyle{plainnat}

\end{document}